\documentclass[aps,prl,twocolumn,groupedaddress,showpacs,amsmath,amssymb,floatfix,superscriptaddress]{revtex4-1}
\usepackage{multirow}
\usepackage{graphicx}
\usepackage{times}
\usepackage{color}
\usepackage{natbib}
\usepackage{array}
\newcolumntype{C}[1]{>{\centering\let\newline\\\arraybackslash\hspace{0pt}}m{#1}}

\newcommand{\ra}{\rightarrow}
\newcommand{\be}{\begin{equation}}
\newcommand{\ee}{\end{equation}}
\newcommand{\bea}{\begin{eqnarray}}
\newcommand{\eea}{\end{eqnarray}}
\newcommand{\etal}{\it {et al.}}

\begin{document}

\title{Detecting Dark Photons with Reactor Neutrino Experiments}

\newcommand{\cupibs}{\affiliation{Center for Underground Physics, Institute for Basic Science, Daejeon 34047, Korea}}
\newcommand{\ust}{\affiliation{University of Science and Technology,
Daejeon 34113, Korea}}
\author{H.K.~Park}\cupibs\ust


\begin{abstract}
We propose to search for light $U(1)$ dark photons, $A'$, produced via kinetically mixing with
ordinary photons via the Compton-like process, $\gamma e^- \rightarrow A' e^-$, in a nuclear
reactor and detected by their interactions with the material in the active volumes of reactor neutrino
experiments.
We derive 95\% confidence-level upper limits on $\epsilon$, the $A'$-$\gamma$ mixing parameter, $\epsilon$, for dark-photon
masses below 1~MeV of $\epsilon~< ~1.3\times 10^{-5}$ and $\epsilon~<~2.1\times 10^{-5}$,  
from NEOS and TEXONO experimental data, respectively. 
This study demonstrates the applicability of nuclear reactors as potential sources of intense 
fluxes of low-mass dark photons. 
\end{abstract}


\maketitle

Despite the many remarkable successes of the Standard Model of particle physics (SM) during the
past several decades, many questions still remain. 
While the SM accurately describes interactions between known particles in terms of the
$U(1)_Y \times  SU(2)_L \times SU(3)_C$ gauge group, it does not incorporate gravity or dark matter,
and does not exclude the possibility that there are additional interactions or gauge bosons.
One simple extension of the SM that addresses the dark matter issue
is the addition of an extra Abelian gauge force, $U(1)'$, with a gauge
boson, commonly called a dark photon (DP), that kinetically mixes with the ordinary photons of
the SM, as suggested in Ref.~\cite{holdom}.
After rotating the kinetically mixed fields to the physical fields, the effective
Lagrangian~\cite{feng} for the photon and DP system
with kinetic mixing parameter ($\epsilon$) is given by
\begin{align}\label{eq:Lagrangian}
\mathcal {L}  =&
	-\frac {1}{4} F_{\mu\nu}F^{\mu\nu}
	-\frac {1}{4} F'_{\mu\nu}F'^{\mu\nu} 
	 + \frac{1}{2} m^2_{A'} A'^2  \nonumber  \\
	 & - e (A_\mu + \epsilon A'_{\mu}) J^{\mu} , \nonumber
\end{align}
where $F_{\mu\nu}$ ($F'_{\mu \nu}$)  is the field strength of photon (DP) field $A_{\mu}$ ($A'_{\mu}$),    
$m_{A'}$ is the DP mass, and $J^{\mu}$ is the current of electrically charged matter.

The DP mass can be generated by either the St\"uckelberg~\cite{feldman} or the Higgs mechanism.
When the SM and the DP are embedded in a grand unified theory, one obtains the kinetic
mixing-parameter at the quantum-loop level to be between $10^{-7}$ and  $10^{-3}$~\cite{hamed}. 
In the context of non-perturbative and large-volume compactifications of string theory constructions,
$\epsilon$  is  estimated to be in the range from $10^{-12}$ to  $10^{-3}$~\cite{abel}.  

If the DP mass is  larger than twice the mass of electron ($2m_e$), it can decay into
an electron-positron pair.  Upper limits on $\epsilon$ for  $m_{A'}~>~2~m_e$
established by electron-positron and hadron colliders, and electron and proton beam-dump experiments are summarized
in Ref.~\cite{acc}.  Constraints on $\epsilon$ for the case where the DP mass is below 1 MeV
come from non-accelerator experiments, including: cosmic microwave background spectrum~\cite{cmb}; 
broadband radio spectra of compact radio sources~\cite{radio}; tests of Coulomb's law~\cite{coul}; 
light-shining-through-wall experiments~\cite{lst}, solar energy loss~\cite{sol}  
helioscope experiments~\cite{heli};  and direct dark matter search experiments~\cite{dm}. 

In antineutrino-electron ($\bar{\nu}_e$-e) scattering experiments that use nuclear reactors
as the $\bar{\nu}_e$ source, constraints on the DP mass and the mixing parameter $\epsilon$
can be established by considering the possibility that DP interactions in the active
volume of the neutrino detector can contribute to  $\bar{\nu}_e$-e scattering signal
as described in Ref.~\cite{nue}.  In this letter, we discuss the possibility that 
reactor neutrino experiments can be exploited to provide a sensitive probe for DPs with masses below 1 MeV. 

Gamma rays of a few MeV produced in a reactor that scatter off electrons in 
the materials of the reactor core can produce DPs via the Compton-like process, 
$\gamma e^- \rightarrow A' e^-$. 
The number of DPs, $N_{A'}$, with the recoil energy $E_{A'}$ from the reactor is given by
the relation 
\be\label{eq:prod}
\frac{dN_{A'}}{dE_{A'}} = 
\int \frac{1}{\sigma_{tot}} \frac{d \sigma_{\gamma \ra A'}}{dE_{A'}} \frac{dN_\gamma}{dE_\gamma} dE_\gamma,
\ee
where $\sigma_{\gamma \ra A'}$ is the cross section for the process
$\gamma e^- \rightarrow A' e^-$, $\sigma_{tot}$ is the total
cross section for photon interacting with material at the gamma energy of $E_\gamma$, and 
$\frac{dN_\gamma}{dE_\gamma}$ is the flux of $\gamma$-rays with energies  
between $E_\gamma$ and $E_\gamma~+~dE_\gamma$.
The cross section for $\sigma_{\gamma \ra A' }$ is given in Ref.~\cite{xsec1}, and,  
in the limit $m_{A'}~\ll~m_e$,  the differential cross section for $\sigma_{\gamma \ra A'}$ can be
expressed as
\be\label{eq:prod}
\frac{d\sigma_{\gamma \rightarrow A' }} {dE_{A'}} \approx
 \epsilon^2 (1 + {\mathcal O}(m_{A'}^2/m_e^2)) \frac{d\sigma_{C}}{d E_r}\Big{|}_{{E_r}=E_{A'}},
\ee
where $\sigma_{C}$ and $E_r$ are
the cross section and the energy of the Compton-scattered $\gamma$-ray, respectively. 

For $\gamma$-ray energies below 1 MeV, DPs are produced with energy $E_{A'}$ less than 1 MeV,
which would be difficult to detect in most reactor neutrino experiments even if they deposit all
of their energies in the detector,  because of large
low-energy backgrounds.  For this reason, the  present study only considers $\gamma$-ray and DP
energies above 1 MeV.  For photons with energies of a few MeV, Compton scattering is the most
important interaction process, dominating over photoelectric absorption and electron-positron
pair production, even for high-atomic-number materials such as uranium. Therefore, it is a reasonable
approximation to use the Compton scattering cross section $\sigma_{C}$ as the total cross section,
$\sigma_{tot}$, for these energies.

 \begin{figure}
\centering
\includegraphics[width=0.48\textwidth]{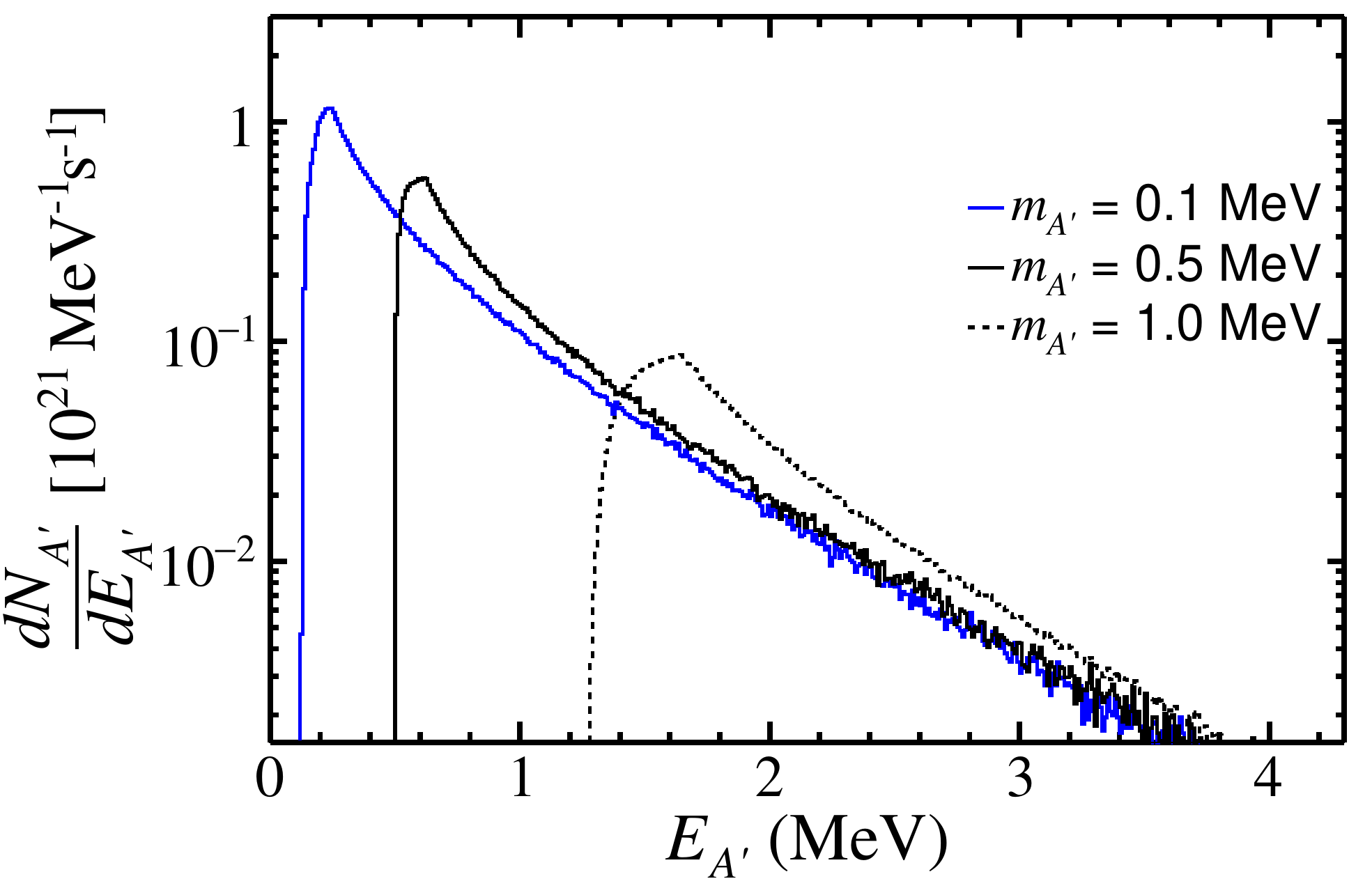}
\caption{The number of DPs per second for a kinetic mixing parameter $\epsilon=1$
emitted at the center of reactor  with thermal power of 1 GW.
Here it is assumed that the reactor is a  point source. 
The blue line, black line and black-dotted lines are for DP masses of 0.1~MeV, 0.5~MeV and 1.0~MeV,
respectively. }
\label{fig:spec}
\end{figure}

Gamma rays are produced inside a nuclear reactor by several different processes: 
emission of prompt $\gamma$-rays in fissions of
$^{235}\rm U$, $^{238} \rm U$, $^{239} \rm Pu$ and $^{241} \rm Pu$  nuclear fuel isotopes; 
$\gamma$ emission from neutron capture and inelastic neutron scattering in the moderator,
fuel and other reactor core materials;
and $\gamma$ emission from the radiative de-excitation of fission daughter nuclei. 
The measured number of prompt $\gamma$-rays  per fission ranges between 6.70 and 7.80 for 
$^{235}\rm U$, $^{239} \rm Pu$   and $^{252} \rm Cf$ nuclear fuel isotopes~\cite{gamma}, which
translates into about $2 \times 10^{20}$ $\gamma$-rays per second with energies below 10~MeV in 
a 1~GW thermal-power reactor.
Since the $\gamma$-ray energy spectrum depends on the fuel composition,
the materials in the core, the core geometry, etc., it is almost impossible to determine an accurate
spectrum for any specific reactor. 
In this study, we  use the $\gamma$-ray flux determined for the FRJ-1 reactor core for
$E_\gamma~\gtrsim~200~\rm keV$~\cite{frj}
\be\label{eq:gray}
\frac{dN_\gamma}{dE_\gamma} = 0.58 \times 10^{18} (\frac{P}{\rm MW}) {\rm exp}[- \frac{E_\gamma}{0.91 \rm MeV}].
\ee
This spectrum was used in the analysis of an axion search experiment performed at the
Bugey nuclear reactor~\cite{bugey}.  
For a reactor with thermal power of 1 GW, Eq.~(\ref{eq:gray})  implies 
$1.76 \times 10^{20}~\gamma$-rays per second for $\gamma$-ray energies above 1 MeV;
 the number of prompt $\gamma$-rays in fissions from the fuel elements
is $6.82 \times 10^{19}~\gamma$-rays per second.  
Although these two estimates differ by a factor of 2.6, their difference does not introduce a
large uncertainty  on the kinetic mixing $\epsilon$ constraint, as discussed below.

Figure~\ref{fig:spec} shows the number of the DPs that would be produced per second 
at  the center of a 1~GW thermal-power reactor as determined using the $\gamma$-ray
spectrum given in Eq.~(\ref{eq:gray}) with the kinetic mixing parameter set at $\epsilon=1$.
In this determination, the reactor is treated as a point source.
The emitted DP flux ($\frac{dN_{A'}}{dE_{A'}}$) for $m_{A'}~<~0.1$ MeV  is not much
different from that for  $m_{A'}=0.1$ MeV shown as the blue curve in Fig.~\ref{fig:spec}.

We consider an $A'$ search for  $m_{A'}~<~1~\rm MeV$. In this mass range, 
the DP can decays to three photons with a decay width given by Ref.~\cite{dec}
\be\label{eq:dec}
\Gamma_{A' \ra 3 \gamma} \approx 2.16 \times 10^{-16} e^4 \epsilon^2 \frac{m_{A'}^9}{m_e^8}.
\ee
This corresponds to a DP decay length ($L_{A'}$) of
\be\label{eq:decl}
L_{A'}= 5.05 \times 10^2 \epsilon^{-2} (\frac{\rm MeV}{m_{A'}})^{10} (\frac{E_{A'}}{\rm MeV})~\rm m.
\ee
Since baselines from reactor to detector for short-baseline reactor neutrino experiments are typically
less than 30~m and the kinetic mixing parameter $\epsilon$ is expected to be much less than
${\mathcal O}(1)$, essentially all of the produced $A'$'s would arrive at detectors without decaying.  
 
The $A'$ can be detected via the DP absorption process, $A' e^- \rightarrow \gamma e^-$.
The cross section for that process, $\sigma_{A' \rightarrow \gamma}$, is given in Ref.~\cite{xsec2} and  
for $m_{A'}~\ll~m_e$ , the differential cross section with respect to the recoil $\gamma$-ray energy can be written as
\be\label{eq:det}
\frac {d \sigma_{A' \rightarrow \gamma}}{d E_{r}} \approx 
\frac{2}{3}\epsilon^2 (1 + {\mathcal O}(m_{A'}^2/m_e^2)) \frac {d\sigma_{C}}{d E_{r}}.
\ee
The total number of observed DP events ($N_{obs}$) from the DP absorption process in a
reactor neutrino experiment would be
\begin{align}\label{eq:sig}
N_{obs}  =& \frac{N_e T }{4\pi R^2}  \int_{E_{A'_1}}^{E_{A'_2}} dE_{A'} \int_{E_{r_1}}^{E_{r_2}}  dE_{r} 
\frac{dN_{A'}}{dE_{A'}} \frac {d \sigma_{A' \rightarrow \gamma}}{d E_r},   
\end{align}
where $T$ is the data taking period,  $N_e$ is the total number of electrons in detector's
fiducial volume,
and $R$ is the distance between the center of reactor  and the detector.  
The  $E_{r_1}$ and $E_{r_2}$ integration limits, $\frac{2 m_e E_{A'}}{m_e + 2 E_{A'}}$
and $E_{A'}$ for $m_{A'}~\ll~m_e$, respectively, are functions of $m_{A'}$ and $E_{A'}$.
The number of $A'$ absorption events are proportional to $\epsilon^4~\sigma_C$.

To extract 95~\% confidence level (C.L.) upper limits on $\epsilon$ 
as a function of the DP mass based on Eq.~(\ref{eq:sig}), 
we take 1.96~times the uncertainty ($\sigma$) of the number of observed $e-\gamma$ events as the number of
upper limit on the number of DP-induced events in the data. 
In this study, we consider the TEXONO~\cite{texo} and NEOS~\cite{neos}
short-baseline reactor experiments.
Both experiments  have similar baselines, reactor power and data taking periods, while
the detector materials, masses and detection energy windows
for the two experiments are different. 
 
\begin{figure*}
\centering
\includegraphics[width=0.9\textwidth]{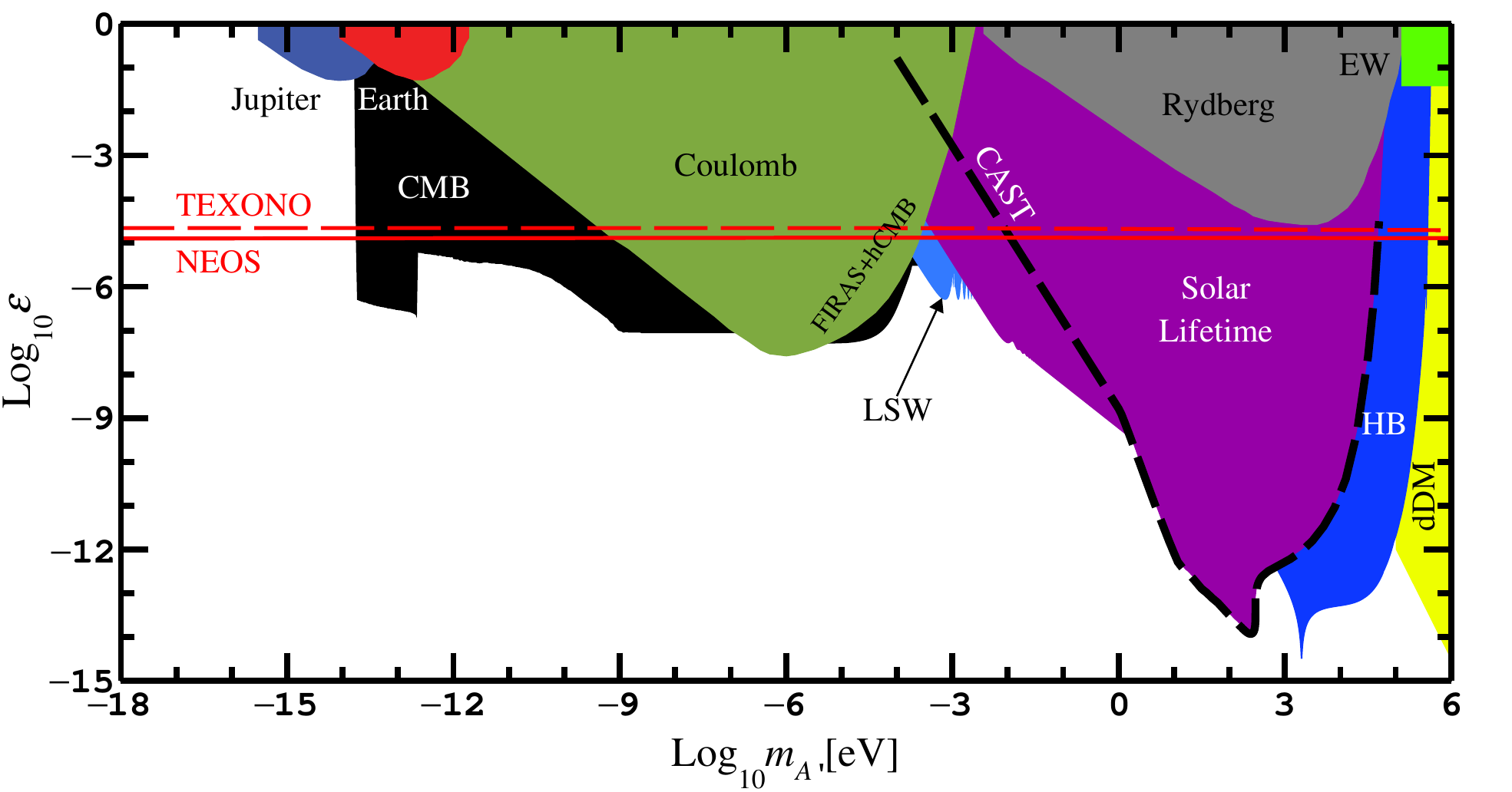}
\caption{Summary of constraints on the DP mass, $m_{A'}$,  {\em versus} the kinetic mixing parameter $\epsilon$. 
Colored regions are excluded regions from astronomical observations, cosmological arguments and experiments. 
A compilation of the constraints and a detailed explanation of each label are given in Ref.~\cite{comp}.
The thick- and dotted-red lines are 95\% CL exclusion upper limit based the NEOS and TEXONO experiments, respectively.}
\label{fig:limit}
\end{figure*}

The TEXONO experiment measured
the $\bar{\nu}_e-e^-$ scattering cross section with a total mass of 187 kg CsI(Tl) scintillating crystal
detector, where the detector is located at a 
distance of 28~m from the core of a 2.9~GW thermal-power reactor. 
The experiment extracts the total number of $\bar{\nu}_e-e^-$ scattering events in the recoil electron energy
3~MeV to 8~MeV to be $[414 \pm 100.6]$,  where the error includes both statistical and systematical uncertainties, for a
160-day  data-taking period; this is consistent with SM expectations for the  number of $\bar{\nu}_e-e^-$
scattering events.  From the  uncertainty, we infer a 95\% CL upper-limit on the number
of DP-induced events of 197.2 and translate that into an
upper limit on $\epsilon$ using Eq.~(\ref{eq:sig}). For this limit determination,
the energies deposited  in the detector by both the recoil $e^-$ and the $\gamma$-ray that is produced in the absorption process
is required to be in the TEXONO experimental limits (between 3~MeV and 8~MeV) by setting
 the integration limits $E_{A'_1}$ at  3~MeV and $E_{A'_2}$ at 8~MeV. The resulting limit is
$\epsilon~<~2.1\times 10^{-5}$ for $m_{A'}~<~1~\rm MeV$ at 95~\% C.L. upper limit. 

The NEOS experiment was a search for sterile neutrino using a 1008~L volume of 
liquid scintillator (LS) detector  located at a distance of 24~m from the center 
of the core of a 2.8~GW thermal-power reactor.  
The experiment took data for 180~days with the reactor on and for 46 days with the reactor off. 
During the reactor-on period, the total number of $e^-/\gamma$ events in the 1~MeV to 5~MeV energy 
range after vetoing cosmic-ray muons was $7.2 \times 10^8$~\cite{neos_priv}, and consistent with the
the background rate determined from the reactor-off data. We, therefore, assume that all of the
reactor-on event candidates are due to background, and use 52,600 events  
(1.96 $\sigma$ of statistical uncertainty of those events) as the 95\% confidence level upper limit 
on the number of observed DP events. 
Setting the integration limits $E_{A'_1}$ to be 1~MeV and $E_{A'_2}$ to be 5 MeV in Eq.~(\ref{eq:sig}), we
determine $\epsilon~<~ 1.3\times 10^{-5}$ for $m_{A'}~<~1~\rm MeV$ at 95~\% C.L. upper limit.

Since  the parameter $\epsilon$ is
inversely proportional to forth root of the $\gamma$-ray spectrum,
the limits for the parameter $\epsilon$ obtained  with the $\gamma$-ray spectrum in Eq.~(\ref{eq:gray}) 
does not introduce a large uncertainty in these upper limits. 
The limits given above are based on Eq.~(\ref{eq:gray});  
using a $\gamma$-ray flux for prompt fission-process $\gamma$-rays
would increase the upper-limits on $\epsilon$ by 30\%.
Since both $\gamma$-ray flux estimations do not correctly include $\gamma$-ray contributions from neutron capture, 
inelastic neutron scattering and other $\gamma$-ray sources in a reactor core, 
the limits  for the parameter $\epsilon$ in our study
would be upper bound estimates.

The  experimental bounds on $\epsilon$ could be substantially improved with better background rejection.
In the NEOS experiment, the $e^-/\gamma$ background events mainly come from ambient $\gamma$ rays and
internal radioactive $^{40} \rm K$ and $^{208} \rm Tl$ contaminations that produce 1.461~MeV and 2.614~MeV
$\gamma$ rays, respectively.  The rejection of these $\gamma$ rays is difficult in the NEOS experiment
because it is a homogeneous LS detector with no segmentation. In comparison, the DANSS detector~\cite{danss}
consists of a similar 1 $\rm m^3$ volume of highly segmented plastic scintillator, that could have potentially
reject ambient background $\gamma$ rays by imposing fiducial cuts.  Internal radioactive backgrounds are
reduced by tight constraints on the intrinsic radiopurity of  the detector materials. Moreover, the DANSS
detector baseline is smaller,  between 9.7 m and 12.2 m from the reactor, and the thermal-power of the
reactor is 3 $\rm GW$. With these improvements, the DANSS experiment can be expected to reach an $\epsilon$
sensitivity level of $10^{-6}$. 

In summary, we propose to search for light DPs produced via the Compton-like process,
$\gamma e^- \rightarrow A' e^-$, in a nuclear reactor core, and detect them via inverse Compton-like
scattering, $A' e^- \rightarrow \gamma e^-$, in a short-baseline-reactor-neutrino detector.
We derived constraints on the kinetic mixing parameter $\epsilon$ for  the NEOS and TEXONO short-baseline
reactor neutrino experiment results, setting 95\% C.L. upper limits of
$\epsilon~<~1.3\times 10^{-5}$ and $\epsilon~<~ 2.1\times 10^{-5}$ for $m_{A'}~<~1~\rm MeV$, respectively.

This work was supported by IBS-R016-D1.  
The author is indebted to: Hye-Sung Lee and Patrick deNiverville for helpful discussions,
Joerg Jaeckel for providing a summary of the DP constraint plot; and Yoomin Oh for providing
information about NEOS and other ongoing experiments.

\end{document}